\documentclass[10pt]{article}

\usepackage{amsmath}
\usepackage{amssymb}
\usepackage{amsthm}
\usepackage{graphicx}
\usepackage{cite}
\usepackage{palatino}
\usepackage[font=footnotesize]{caption}
\usepackage{subcaption}
\usepackage{float}
\usepackage{hyperref}

\begin{document}

\title{Commutative Algebra Learning for Protein Flexibility Analysis}
	\author{Honghao Zhang$^1$ and Hongsong Feng$^{1}$\footnote{
		Corresponding author.		Email: hfeng2@charlotte.edu} \\
	\\
	$^1$ Department of Mathematics and Statistics,\\
	University of North Carolina at Charlotte, Charlotte, NC 28223, USA\\
}
\date{\today}

\maketitle

\begin{abstract}
Protein flexibility, commonly quantified by B-factors, is closely related to protein structure and function. However, accurate B-factor prediction remains challenging due to the multiscale nature of protein structures and the complexity of atomic interactions. In this work, we propose a commutative algebra-based learning framework, termed CAL, for protein B-factor prediction. Unlike many biomolecular prediction tasks that rely primarily on global structural representations, B-factor prediction requires an accurate characterization of the local geometric environments surrounding individual atoms. To address this challenge, CAL employs commutative algebra theory to construct localized algebraic descriptors at multiple spatial scales. On a benchmark dataset of 364 proteins, CAL improves prediction accuracy by 34.5\% over the classical Gaussian network model (GNM). Extensive experiments demonstrate that CAL achieves robust and consistent performance across diverse datasets and is competitive with existing state-of-the-art methods. Furthermore, by integrating CAL with machine learning, we develop a blind prediction model capable of cross-protein B-factor prediction. Overall, CAL provides an effective, efficient, and mathematically principled framework for protein flexibility prediction and offers a powerful approach for analyzing and predicting localized structural properties in complex biomolecular systems.
\end{abstract}

\section{Introduction}

Life processes critically depend on proteins, which play central roles in cellular signaling, genetic regulation, transcription and translation, as well as protein–ligand interactions \cite{petsko2004protein}. It is traditionally believed that protein function is determined by its three-dimensional structure. Although amino acid sequences can undergo substantial variations and chemical modifications, proteins often retain their folding capability and biological functionality \cite{Anfinsen1973}.
However, increasing evidence suggests that proteins are not static structures but instead undergo continuous conformational fluctuations and dynamic transitions across their energy landscapes \cite{frauenfelder1991}. Such structural flexibility is an intrinsic property of proteins and plays a crucial role in their biological functions. In particular, proteins can adapt to different environments through conformational rearrangements, enabling specific interactions with DNA, RNA, ions, cofactors, and various ligands \cite{Hayes2025}.

The B-factor, also known as the temperature factor or atomic displacement parameter, is a fundamental quantity in X-ray crystallography that describes the mean-square displacement of atoms around their equilibrium positions. It provides a statistical measure of atomic fluctuations in protein structures \cite{trueblood1996}.From a structural biology perspective, B-factors reflect local flexibility and conformational variability within protein structures. Recent studies have further demonstrated that B-factors serve as effective structural descriptors for protein flexibility, capturing residue-level heterogeneity in dynamic environments and being widely used in structural and functional analysis \cite{Sun2019}.Therefore, the B-factor can be regarded as an informative structural signal for protein flexibility rather than merely a byproduct of crystallographic refinement, forming the basis for structure-based modeling of protein properties.

Protein structural flexibility can be carried out through molecular dynamics (MD) and normal mode analysis (NMA). MD directly simulates the time evolution of protein systems at the atomic level by numerically integrating Newton’s equations of motion under empirical potential energy functions, thereby providing a detailed description of conformational fluctuations and dynamical behavior \cite{McCammon1977}. Although MD is capable of capturing realistic molecular motions derived from physical force fields, its reliance on very small time steps and long integration times leads to high computational cost, which limits its applicability to large-scale systems.

In contrast, NMA expands the potential energy function around an equilibrium configuration using a second-order Taylor approximation and reformulates the dynamical problem as an eigenvalue decomposition of the Hessian matrix. The eigenvectors describe collective vibrational modes of the protein, while the eigenvalues correspond to their associated frequencies, enabling analytical characterization of low-frequency large-scale motions \cite{Levitt1985}. This framework also allows a direct connection between theoretical fluctuations and experimentally measured B-factors. Moreover, this formulation has been integrated into empirical force-field-based computational frameworks for structural optimization and dynamical analysis \cite{Brooks1983}. However, since it still requires construction and diagonalization of high-dimensional Hessian matrices, its computational cost remains significant at the all-atom level, which has motivated the development of coarse-grained normal mode models that preserve dominant collective motions while substantially reducing computational complexity \cite{Park2013}.

To reduce the computational cost of MD and all atom NMA, elastic network models (ENMs) represent proteins as coarse-grained spring networks composed of interacting residues. The Gaussian network model (GNM) assumes isotropic residue fluctuations and characterizes residue couplings using a Kirchhoff contact matrix, enabling efficient prediction of structural fluctuations and experimental B-factors \cite{Bahar1997}. This framework was subsequently used to analyze collective motions at different scales and revealed that low-frequency modes are often associated with functional motions, whereas high-frequency modes contribute to structural stability \cite{Bahar1998}. However, GNM only describes fluctuation magnitudes and neglects directional information. To address this limitation, the anisotropic network model (ANM) incorporates three-dimensional directional correlations through a Hessian matrix representation, allowing explicit characterization of collective protein motions \cite{Atilgan2001}.

Although elastic network models\cite{Bahar1997,Bahar1998,Atilgan2001} have been widely applied to protein flexibility analysis, their performance typically relies on matrix diagonalization procedures and fixed cutoff parameters, making it difficult to simultaneously capture interactions across multiple characteristic length scales \cite{Opron2015}. To address these limitations, the Flexibility-Rigidity Index (FRI) was proposed to characterize topological connectivity through distance-dependent correlation functions, allowing protein flexibility and B-factors to be predicted directly from structural information without constructing interaction Hamiltonians or performing eigenvalue decomposition \cite{Xia2013Multiscale}. Subsequently, a stochastic framework was introduced to describe protein fluctuations through transition probability processes, providing a non-Hamiltonian perspective for flexibility analysis \cite{Xia2013Stochastic}. Building upon this framework, the fast FRI (fFRI) reduced computational complexity from $O(N^2)$ to $O(N)$ through local neighborhood searching, while anisotropic FRI (aFRI) further incorporated directional information to characterize collective protein motions \cite{Opron2014}. Furthermore, the multiscale FRI (mFRI) combined multiple correlation kernels operating at different characteristic scales to simultaneously capture local and long-range interactions, leading to improved B-factor prediction accuracy and enhanced applicability to large and complex macromolecular systems \cite{Opron2015}.

In recent years, protein flexibility analysis has progressively evolved from traditional physics-based modeling toward structure-driven mathematical learning frameworks. In this direction, the Persistent Sheaf Laplacian (PSL) method employs spectral sheaf structures to jointly capture local connectivity and global topological information, leading to improved B-factor prediction performance \cite{Hayes2025}; the Multiscale Differential Geometry (mDG) framework introduces curvature-based representations on low-dimensional manifolds to characterize multiscale geometric variations of proteins, thereby enhancing expressive power for complex structures \cite{Feng2025}; while the Magnitude approach constructs structural invariants from finite metric spaces, representing proteins as weighted geometric objects and extracting global topological features, achieving strong performance in flexibility prediction tasks \cite{Bi2025}. Although these methods arise from different mathematical formalisms, they collectively indicate a shift from explicit physics-based modeling toward topology and geometry driven representation frameworks.

Commutative algebra is a branch of mathematics that studies commutative rings, ideals, modules, and their associated algebraic structures \cite{suwayyid2026persistent}. In recent years, there has been growing interest in applying topological data analysis (TDA) to data science and machine learning through a variety of techniques from algebraic topology \cite{wang2020persistent,shen2024persistent,liu2026local,memoli2022persistent}. These approaches have achieved remarkable success, particularly in biomolecular modeling and drug discovery \cite{shen2024persistent,feng2023virtual,liu2026persistent}. Inspired by the success of TDA, commutative algebra, one of the mathematical foundations of algebraic geometry and homological algebra, has recently emerged as a powerful framework for characterizing structural relationships and algebraic invariants in complex systems. By encoding rich algebraic structures, commutative algebra provides information complementary to geometric and topological descriptors, leading to more interpretable, mathematically rigorous, and predictive models. These advantages have enabled successful applications in biomolecular modeling, genomics, and materials science \cite{Feng2025CAML,zia2025cap,wee2025canet,suwayyid2025cakl,khaemba2026commutative}.

The aim of this work is to develop a novel model for protein flexibility analysis based on our recently proposed commutative algebra framework \cite{suwayyid2026persistent,Feng2025CAML}. Unlike many biomolecular prediction tasks that rely primarily on global structural representations, protein flexibility (B-factor) prediction requires accurate characterization of localized geometric environments surrounding individual atoms. To address this challenge, we extend our commutative algebra theory to construct localized algebraic descriptors specifically tailored for protein flexibility analysis. The remainder of this manuscript is organized as follows. \autoref{sec:method} presents the proposed methodology and describes how the commutative algebra framework is employed to construct atomic-level features for B-factor prediction. \autoref{sec:results} reports the experimental results and evaluates the performance of the proposed models. Finally, \autoref{sec:conclusion} concludes the paper with a summary of our novel theory for modeling protein flexibility.

\section{Theory and Methods} \label{sec:method}

\begin{figure}[H]
	\centering
	\includegraphics[width=1.0\linewidth]{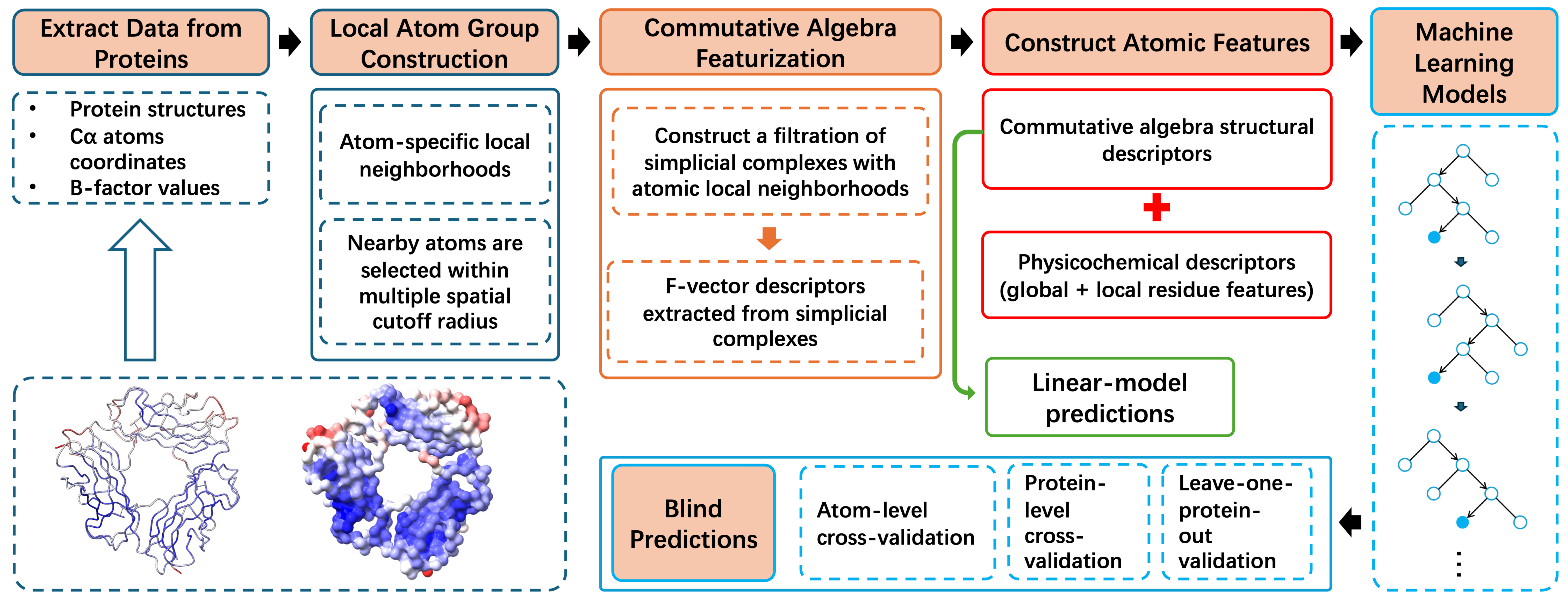}
	\caption{
		This figure illustrates the overall pipeline of the proposed CAL-based framework for protein B-factor prediction, including data input, local structure construction, CAL feature generation based on multiscale topology, integration with additional structural descriptors, and prediction using both linear and ensemble machine learning models under multiple validation settings.
	}
	\label{fig:concept}
\end{figure}

Protein flexibility is fundamentally determined by molecular interactions occurring across multiple characteristic length scales, ranging from local covalent and hydrogen-bond interactions to long-range noncovalent effects. As a consequence, effective flexibility models must be able to capture structural information over a broad range of spatial scales. Recent studies have demonstrated that geometric and topological representations provide powerful tools for extracting such multiscale information directly from protein structures. Since protein structures can be naturally represented as three-dimensional point clouds formed by atomic coordinates, combinatorial topological constructions offer a mathematically rigorous framework for describing the relationships among neighboring atoms and their evolution across scales. In this work, we employ a simplicial-complex-based filtration framework to characterize multiscale structural organization and construct descriptors for protein flexibility prediction\cite{Opron2015,Feng2025,Hayes2025}. \autoref{fig:concept} demonstrates how we develop models using our commutative algebra theory for protein flexibility analysis and prediction. 

\subsection{Simplicial Complexes and Filtration}

Simplicial complexes provide a combinatorial framework for describing the geometric organization of point-cloud data. The fundamental building block of a simplicial complex is a simplex. Define $\sigma=[v_0,v_1,\ldots,v_k],s$ where $v_0,v_1,\ldots,v_k$ are $k+1$ affinely independent points. The convex hull generated by these vertices is called a $k$-simplex. Typical examples include a vertex (0-simplex), an edge (1-simplex), a triangle (2-simplex), and a tetrahedron (3-simplex). Any simplex generated by a subset of the vertices is referred to as a face of the simplex.

A simplicial complex is a collection of simplices satisfying the closure property. Specifically, if a simplex $\sigma$ belongs to a simplicial complex $K$, then all faces of $\sigma$ must also belong to $K$. In addition, the intersection of any two simplices is either empty or a common face of both simplices. The dimension of a simplicial complex is defined as the highest dimension among all simplices contained in the complex.

To construct simplicial complexes from point-cloud data, we employ the Vietoris--Rips (VR) complex. Let
$X={x_1,x_2,\ldots,x_n},$ where $x_i\in\mathbb{R}^3$. For a prescribed distance parameter $\varepsilon$, a simplex is included whenever all pairwise Euclidean distances among its vertices do not exceed $\varepsilon$. The resulting VR complex is defined as
\begin{equation}
VR_{\varepsilon}(X)
=
\left\{
\sigma\subseteq X :
d(x_i,x_j)\le\varepsilon,
\ \forall x_i,x_j\in\sigma
\right\}
\label{eq:vr}
\end{equation}
where $d(x_i,x_j)$ denotes the Euclidean distance between points $x_i$ and $x_j$.

As the distance parameter increases, additional edges, triangles, and higher-dimensional simplices are progressively incorporated into the complex, yielding a nested sequence
\begin{equation}
VR_{\varepsilon_1}(X)
\subseteq
VR_{\varepsilon_2}(X)
\subseteq
\cdots
\subseteq
VR_{\varepsilon_m}(X),
\qquad
\varepsilon_1<\varepsilon_2<\cdots<\varepsilon_m,
\label{eq:filtration}
\end{equation}
which is referred to as a filtration. Filtration records the evolution of combinatorial structures across increasing spatial scales and provides the mathematical foundation for the multiscale feature representation developed in this work.

\subsection{Persistent Facet Ideals}

In commutative algebra, the combinatorial structure of a simplicial complex can be characterized through its facets, namely the maximal simplices under inclusion. Let $K$ be a finite simplicial complex on the vertex set
\begin{equation*}
V=\{v_1,v_2,\dots,v_n\}.
\end{equation*}
A simplex $\sigma \in K$ is called a facet if it is not contained in any larger simplex in $K$, that is,
\begin{equation*}
\sigma \subseteq \tau,\ \tau \in K \Rightarrow \sigma=\tau.
\end{equation*}
The collection of all facets of $K$ is denoted by
\begin{equation*}
\mathcal{F}(K)=\{\sigma \in K \mid \nexists \tau \in K \text{ such that } \sigma \subsetneq \tau\}.
\end{equation*}
For each facet
\[
\sigma=\{v_{i_1},v_{i_2},\dots,v_{i_r}\},
\]
one associates the square-free monomial
\[
m_\sigma=x_{i_1}x_{i_2}\cdots x_{i_r}
\]
in the polynomial ring
\begin{equation*}
R=k[x_1,x_2,\dots,x_n].
\end{equation*}
The facet ideal of $K$ is then defined as the monomial ideal generated by all facet monomials,
\begin{equation*}
I_F(K)=\langle m_\sigma \mid \sigma \in \mathcal{F}(K)\rangle.
\end{equation*}
Unlike simplices, facets do not evolve monotonically under filtration. As the filtration parameter increases, higher-dimensional simplices may appear and absorb previously maximal simplices, causing them to lose their facet status. Consequently, facets exhibit birth and death events analogous to those in persistent homology, although the objects being tracked are maximal simplices rather than homology classes.

Consider a Vietoris--Rips filtration
\begin{equation*}
\emptyset=K_{\varepsilon_1}\subseteq K_{\varepsilon_2}\subseteq \cdots \subseteq K_{\varepsilon_m}.
\end{equation*}
Let
\begin{equation*}
\mathcal{P}_k^\varepsilon=
\{P_\sigma \mid \sigma \in \mathcal{F}(K_\varepsilon),\ \dim(\sigma)=k\}
\end{equation*}
denote the collection of $k$-dimensional facet ideals at filtration scale $\varepsilon$. For two filtration scales satisfying $\varepsilon \le \varepsilon'$, the persistent $k$-dimensional facet ideals are defined by
\begin{equation*}
\mathcal{P}_k^{\varepsilon,\varepsilon'}=
\mathcal{P}_k^\varepsilon \cap \mathcal{P}_k^{\varepsilon'}.
\end{equation*}
The corresponding facet persistence number is
\begin{equation*}
\beta_k^{\varepsilon,\varepsilon'}=
\left|\mathcal{P}_k^{\varepsilon,\varepsilon'}\right|,
\end{equation*}
which records the number of $k$-dimensional facet ideals surviving simultaneously at both filtration scales. This construction provides an algebraic analogue of persistence barcodes for maximal simplices and characterizes the evolution of combinatorial structures along the filtration process. One illustration of persistent facet ideals is given in \autoref{fig:vr_fvector}c. Although persistent facet ideals contain rich algebraic information, explicit facet tracking and ideal construction become computationally expensive for large biomolecular complexes due to the combinatorial growth of high-dimensional simplices. Motivated by persistent facet evolution while aiming to maintain computational efficiency, we adopt a coarse-grained combinatorial representation based on simplex counting, namely the multiscale $f$-vector representation introduced in the next section. Our previous investigation on applying commutative algebra method shows the effectiveness of facet ideal and $f-$ vector in diverse biomolecular systems \cite{zia2025cap,wee2025canet,suwayyid2025cakl,khaemba2026commutative,Feng2025CAML}. Persistent facet ideal is effective in biomolecular modeling, while another index called f-vector which has equivalence but is more computationally efficient is also useful. We mainly employ f-vector to characterize atoms for B-factor analysis.

\subsection{F-vector Representation}

To quantitatively characterize the combinatorial structures contained in a simplicial complex, we employ the $f$-vector, a descriptor that records the numbers of simplices in different dimensions.

Let $K$ be a simplicial complex with maximal dimension $d$. The $f$-vector of $K$ is defined as
\begin{equation}
f(K)
=
(f_0,f_1,\ldots,f_d),
\label{eq:fvector}
\end{equation}
where $f_k$ denotes the number of $k$-dimensional simplices contained in $K$. In particular, $f_0$, $f_1$, and $f_2$ represent the numbers of vertices, edges, and triangles, respectively.

For the Vietoris--Rips filtration introduced in Section 2.1, each filtration scale $\varepsilon$ determines a simplicial complex
\begin{equation}
K(\varepsilon)
=
VR_{\varepsilon}(X).
\label{eq:Keps}
\end{equation}

Accordingly, an $f$-vector can be associated with every filtration scale,
\[
f(K(\varepsilon))
=
\bigl(
f_0(\varepsilon),
f_1(\varepsilon),
\ldots,
f_d(\varepsilon)
\bigr),
\label{eq:fvector_filtration}
\]
where
\begin{equation}
f_k(\varepsilon)
=
\#\{
\sigma \in K(\varepsilon):
\dim(\sigma)=k
\},
\label{eq:fk}
\end{equation}
and $\#$ denotes the cardinality of a set.

As the filtration parameter $\varepsilon$ increases, new simplices are progressively incorporated into the complex. Consequently, the simplex counts evolve across filtration scales, producing a family of functions
\[
f_k(\varepsilon),
\qquad
k=0,1,\ldots,d,
\label{eq:fcurve}
\]
which are referred to as \emph{$f$-vector curves}. These curves describe the evolution of combinatorial structures throughout the filtration process and provide multiscale information about the underlying point cloud.

\begin{figure}[H]
    \centering
    \includegraphics[width=0.75\linewidth]{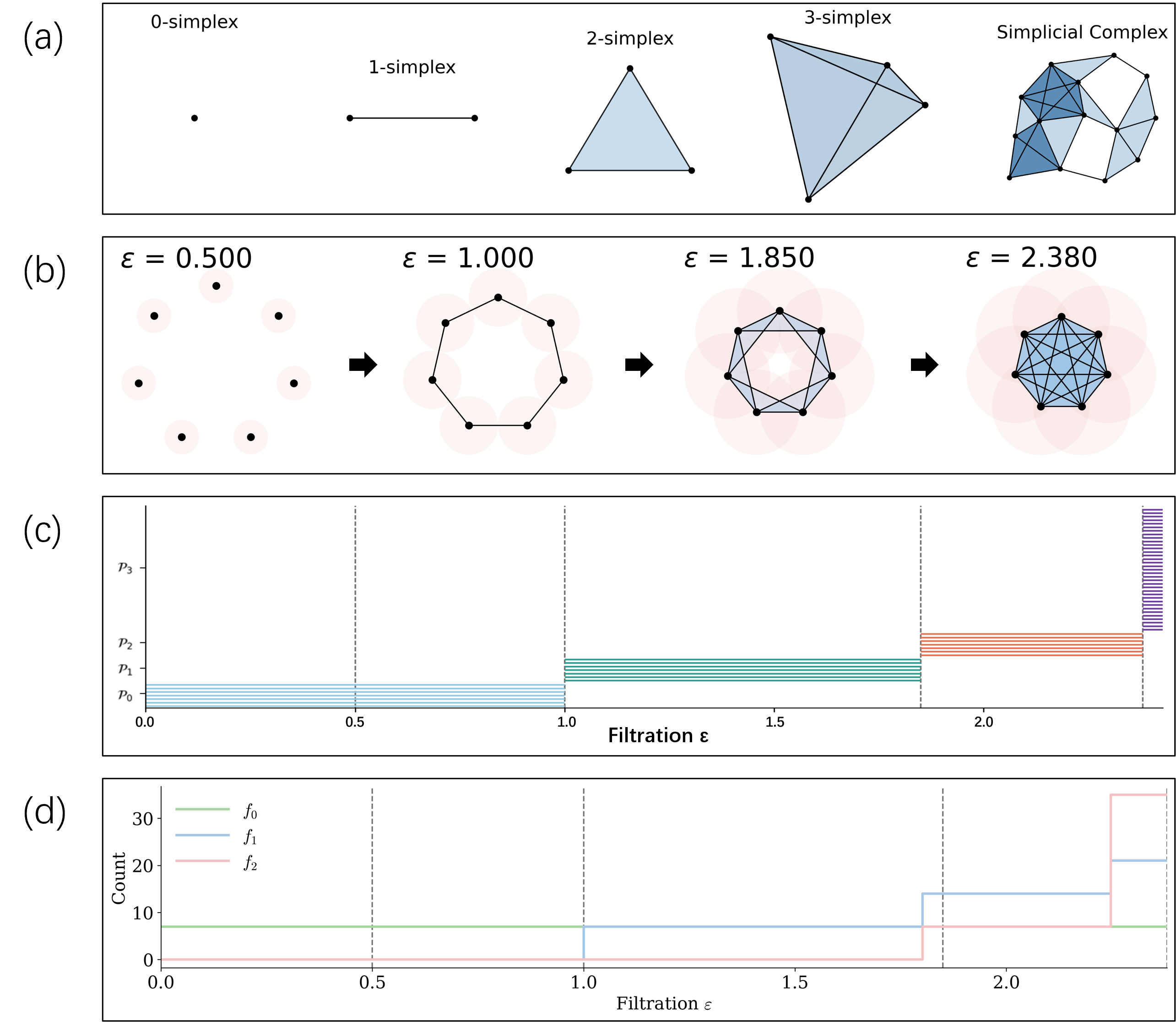}
    \caption{
    Construction of commutative algebra-based descriptors.
    (a) Examples of simplices of dimensions $0$--$3$, including a vertex, an edge, a triangle, and a tetrahedron, together with an example simplicial complex composed of simplices of different dimensions.
    (b) Evolution of a Vietoris--Rips filtration as the filtration parameter $\varepsilon$ increases. New simplices are progressively introduced when pairwise distances satisfy the filtration criterion, resulting in increasingly complex combinatorial structures.
    (c) Facet-based multiscale representation of simplicial structures under the filtration process. Each horizontal segment represents the persistence interval of simplices across filtration scales, where different rows correspond to simplices of different dimensions $\mathcal{P}_0$–$\mathcal{P}_3$. The visualization captures the birth and disappearance of simplicial structures as the filtration parameter increases, providing a barcode-like representation of their multiscale evolution.
    (d) Corresponding $f$-vector curves showing the simplex counts $f_0$, $f_1$, and $f_2$ as functions of $\varepsilon$. The vertex count remains constant throughout the filtration, whereas the numbers of edges and triangles increase as additional simplices are incorporated into the complex.
    }
    \label{fig:vr_fvector}
\end{figure}

To provide an intuitive illustration of the above construction, ~\autoref{fig:vr_fvector} visualizes the formation of simplicial complexes and their multiscale evolution under Vietoris–Rips filtration. The figure demonstrates how simplices of different dimensions are progressively generated as the filtration parameter increases, leading to increasingly complex combinatorial structures. In addition, the visualization extends across multiple levels, including simplex formation, Vietoris–Rips geometric growth, facet evolution, and $f$-vector statistics. The corresponding changes in simplex counts across scales further motivate the construction of multiscale topological descriptors used in this work.

\subsection{Atom-specific f-vector}

The $f$-vector curves introduced in Section 2.2 provide a multiscale description of the combinatorial structures generated during a filtration. To extract local structural information for protein flexibility analysis, we construct atom-centered topological descriptors from local neighborhoods of individual atoms.

Let $X=\{x_1,x_2,\ldots,x_N\},$ where \(x_i\in\mathbb{R}^3\) denotes the Cartesian coordinate of the \(i\)-th atom. For a central atom \(x_i\), a local neighborhood is defined as
\begin{align*}
\mathcal N_R(x_i)
=
\{x_j\in X :
0 < \|x_j-x_i\| \le R\},
\end{align*}
where \(R\) denotes a prescribed neighborhood radius. The exclusion of the central atom avoids trivial self-interactions, while the neighborhood construction enables the extraction of localized structural information around the atom of interest. Similar atom-centered neighborhood representations have been employed in topological approaches for protein flexibility analysis \cite{Bramer2020}.

For each neighborhood \(\mathcal N_R(x_i)\), a Vietoris--Rips filtration is constructed as described in Section 2.1. Let
\[
0 \le \varepsilon_1 < \varepsilon_2 < \cdots < \varepsilon_m \le R
\]
be a sequence of filtration scales. At each filtration scale \(\varepsilon_\ell\), the corresponding simplicial complex \(K(\varepsilon_\ell)\) generates an $f$-vector
\[
f(K(\varepsilon_\ell))
=
\bigl(
f_0(\varepsilon_\ell),
f_1(\varepsilon_\ell),
\ldots,
f_d(\varepsilon_\ell)
\bigr),
\qquad
\ell=1,\ldots,m.
\]
Very small filtration scales typically contain only a limited number of neighboring interactions and therefore provide little combinatorial information. Consequently, the feature construction focuses on the informative portion of the filtration and retains the simplex-count functions
\[ f_1(\varepsilon)
\quad\text{and}\quad
f_2(\varepsilon),\]
which correspond to the numbers of edges and triangles, respectively. These quantities characterize local connectivity and higher-order combinatorial structures generated during the filtration process.

The sampled values of the selected $f$-vector curves are concatenated to form a finite-dimensional descriptor associated with the central atom,
\begin{equation}
\mathbf{x}_i
=
\Big[
f_1(\varepsilon_1),\ldots,f_1(\varepsilon_m),
f_2(\varepsilon_1),\ldots,f_2(\varepsilon_m)
\Big].
\end{equation}
To capture structural information at multiple spatial scales, descriptors obtained from different local neighborhoods can be further combined through feature concatenation. The resulting vector provides a multiscale combinatorial representation of the local geometric environment surrounding each atom and serves as the fundamental topological descriptor employed in this work.

\subsection{Additional Features for Machine Learning}

In addition to the proposed F-vector topological descriptors, a collection of protein structural descriptors was incorporated as additional features to further improve the prediction performance of the machine learning models. While the proposed topological features characterize the local geometric and topological organization of protein structures, they do not explicitly encode the biochemical properties or conformational information of individual residues\cite{Feng2025}. Therefore, conventional structural descriptors were combined with the proposed topological features to construct a more comprehensive representation of protein flexibility.

The additional features consist of global and local structural descriptors. Global descriptors characterize the overall properties of an entire protein, including crystallographic resolution, crystallographic value $R$, and the total number of heavy atoms. These quantities are directly extracted from the corresponding PDB files. Among them, the resolution and $R$-value reflect the quality of the experimentally determined protein structure, whereas the total number of heavy atoms provides a coarse measurement of protein size\cite{Feng2025}.

The local descriptors characterize the structural environment surrounding each residue. In particular, packing density is used to quantify the compactness of the local neighborhood around each residue and is defined as
\begin{equation}
P_i^{d}=\frac{N_d}{N},
\end{equation}
where $N_d$ denotes the number of heavy atoms located within a prescribed distance threshold $d$ from the $i$-th residue, and $N$ denotes the total number of heavy atoms in the protein. To characterize local structural environments at different spatial scales, three packing density descriptors are computed using short-range ($d<3$ \AA), medium-range ($3\le d<5$ \AA), and long-range ($d\ge5$ \AA) distance intervals, respectively\cite{Feng2025}.

In addition, residue-specific information, including amino acid identity and atomic occupancy, was extracted directly from the PDB files. Amino acid identity provides residue-specific biochemical information, whereas occupancy reflects the experimental occupancy and reliability of each atom in the crystal structure\cite{Feng2025}. 

To further incorporate local conformational information, protein secondary structures were assigned using the STRIDE software. STRIDE is a knowledge-based secondary structure assignment method that combines hydrogen-bond energy with backbone torsion-angle information to identify secondary structural elements directly from protein atomic coordinates. Besides secondary structure assignment, STRIDE also provides the backbone dihedral angles $\phi$ and $\psi$, together with the solvent accessible area for each residue. These descriptors provide complementary biochemical and conformational information that cannot be directly captured by topological representations\cite{Heinig2004}.

Finally, the proposed F-vector topological descriptors were concatenated with the above global and local structural descriptors to construct the final feature representation for the machine learning models. By integrating multiscale topological information with conventional structural descriptors, the resulting feature vector provides a more comprehensive representation of protein flexibility.

\section{Results}\label{sec:results}

\subsection{Data Sets}

In this study, we evaluate the proposed method on two benchmark protein datasets for protein flexibility and B-factor prediction. The first dataset consists of 364 protein structures, obtained from a\cite{Opron2014,Opron2015}. The second dataset is taken from \cite{Park2013} and contains three subsets of proteins categorized by structural size, namely small, medium, and large sets, which include 33, 36, and 35 proteins, respectively. These subsets are constructed from non-redundant representative protein structures and are used for residue-level fluctuation evaluation across different structural scales. These three subsets are all subsets of the first dataset containing 364 proteins.

For the blind prediction experiments, several proteins were excluded from the original dataset to ensure data consistency and feature availability. Proteins 1OB4, 1OB7, 2OXL, and 3MD5 were removed because STRIDE failed to generate the required structural features for these proteins. In addition, proteins 1NKO, 2OCT, and 3FVA were excluded since some of their residues were reported with zero B-factors, which are physically unrealistic for flexibility analysis. Furthermore, proteins 3DWV, 3MGN, 4DPZ, 2J32, 3MEA, 3A0M, 3IVV, 3W4Q, and 2DKO were removed due to inconsistencies between the structural information processed by STRIDE and the corresponding original PDB files. After applying these filtering criteria, a total of 348 proteins remained for the blind prediction experiments.

\subsection{Feature Optimization and Model Evaluation}

To evaluate the predictive capability of the proposed CAL representation, we employ a linear regression model using only the constructed 14-dimensional CAL feature vector. 

As described in Section 2.3, the CAL representation is constructed from multiscale topological signatures computed on residue-centered local neighborhoods. Two interaction scales, denoted as $R_1$ and $R_2$, are used to capture structural information from different spatial ranges. For each scale, topological features are extracted from dimension one and dimension two, and sampled over multiple filtration levels to form scale-specific feature curves. In this work, each interaction scale is further discretized into four proportional sub-scales ${R/4, R/2, 3R/4, R}$, including the original scale itself, to capture progressively increasing neighborhood information in a consistent manner.

The final residue-level feature is obtained by concatenating multiscale descriptors. Specifically, each dimension provides four sampled values across filtration scales. To avoid redundancy at the smaller scale, the first sampled point of the $R_1$ scale is removed. Therefore, the final feature dimensionality is constructed as
\[
(4-1)\times 2 + 4\times 2 = 14.
\]

Prediction performance is evaluated using the Pearson correlation coefficient (PCC), defined as
\begin{equation}
\mathrm{PCC}(x,y)
=
\frac{
\sum_{m=1}^{M}
(B_m^{e} - \bar{B}^{e})
(B_m^{t} - \bar{B}^{t})
}{
\sqrt{
\sum_{m=1}^{M}(B_m^{e} - \bar{B}^{e})^2
}
\sqrt{
\sum_{m=1}^{M}(B_m^{t} - \bar{B}^{t})^2
}
},
\label{eq:fk}
\end{equation}

where $B_m^{e}$ and $B_m^{t}$ denote the experimental and predicted B-factor values of residue $m$, respectively, $\bar{B}^{e}$ and $\bar{B}^{t}$ are their corresponding mean values, and $M$ is the number of residues in the protein.

\begin{figure}[H]
    \centering
    \includegraphics[width=0.92\linewidth]{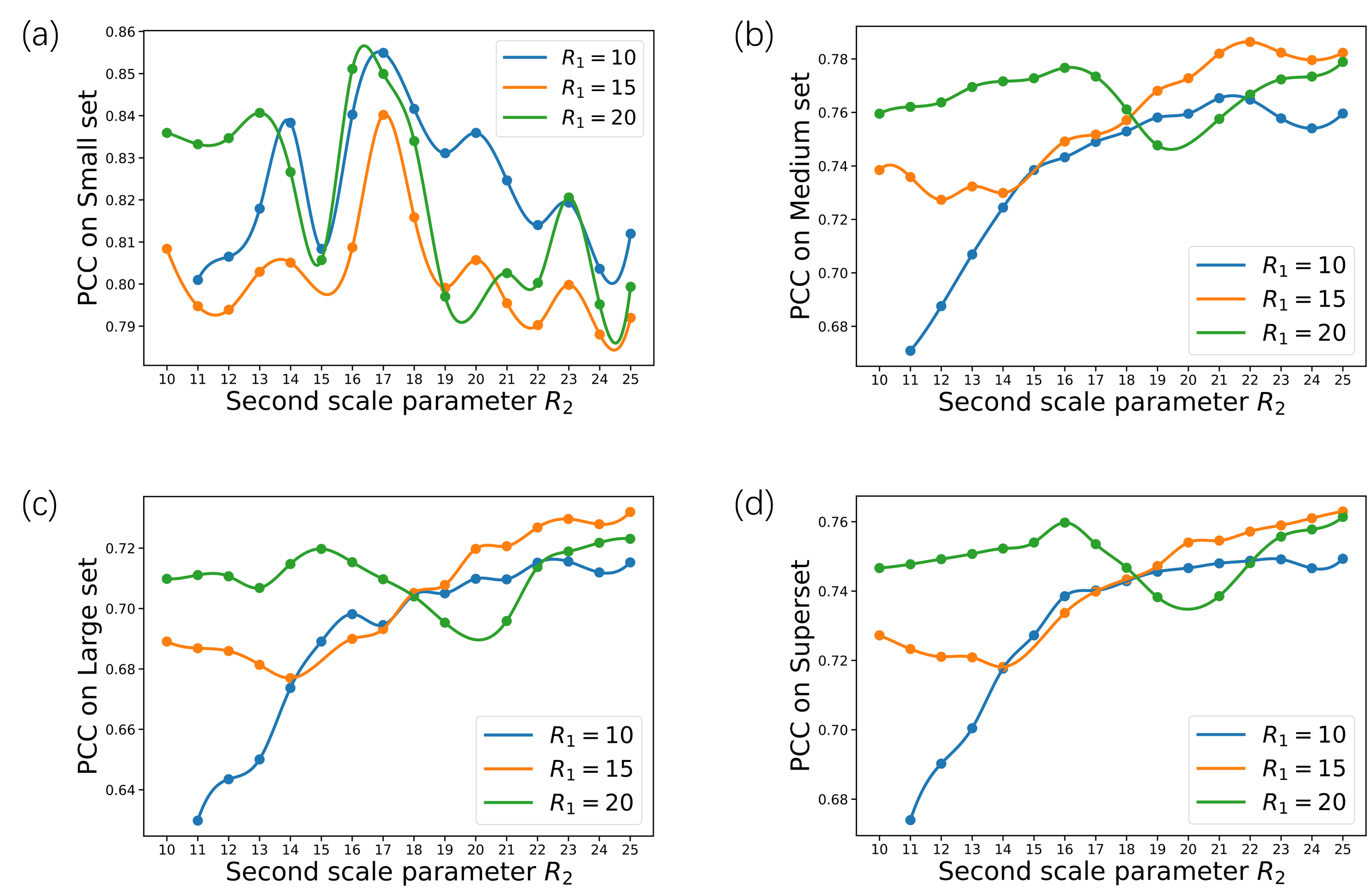}
    \caption{
    Evaluation of the 14-dimensional CAL representation under different parameter settings on four datasets (small, medium, large, and superset). 
    Each panel reports the average PCC of linear regression models as a function of $R_2$ for different $R_1$ values. 
    The results indicate consistent trends across datasets, with the best overall performance observed around $(R_1, R_2) = (20, 16)$.
    }
    \label{fig:pcc}
\end{figure}

To determine optimal parameter settings, we perform a grid search over different combinations of $(R_1, R_2)$ using four benchmark datasets, including small, medium, large, and superset protein sets. The corresponding results are summarized in ~\autoref{fig:pcc}. The results indicate that increasing the secondary scale $R_2$ generally improves predictive performance across all datasets, suggesting the importance of long-range structural interactions. Among all tested configurations, $(R_1, R_2) = (20, 16)$ consistently achieves stable and competitive performance across all four datasets. Therefore, this parameter setting is adopted in all subsequent experiments.

Across all datasets, the linear regression model demonstrates strong and consistent predictive performance using only the 14-dimensional CAL representation, indicating that the learned topological features provide a linearly informative embedding of residue-level flexibility. On the small dataset, the model achieves the best performance, with a mean Pearson correlation of 0.851 and an RMSE of 1.73, along with an $R^2$ of 0.735, suggesting that CAL features are highly effective in capturing local structural fluctuations in compact protein systems. As the dataset scale increases, a gradual performance degradation is observed. For the medium dataset, the Pearson correlation decreases to 0.777 with an RMSE of 2.93 and an $R^2$ of 0.611, while for the large dataset, the performance further drops to 0.715 in Pearson correlation and 3.70 in RMSE, accompanied by an $R^2$ of 0.521, reflecting increased structural heterogeneity and modeling difficulty. Interestingly, the superset dataset yields a moderate recovery in performance, with a Pearson correlation of 0.760 and an RMSE of 3.17, indicating that although cross-protein variability introduces additional noise, the representation remains robust and generalizable across heterogeneous structures. Overall, these results suggest a clear trend of decreasing linear predictability with increasing structural complexity, while the relatively stable performance across datasets highlights the robustness and transferability of the CAL representation.

Building upon the linear regression results and the observed robustness of the CAL representation across datasets, we further evaluate its predictive capability through a systematic comparison with several representative baseline methods, including ASPH (B)\cite{Bramer2020}, ASPH (W)\cite{Bramer2020}, opFRI\cite{Opron2014}, pfFRI\cite{Opron2014}, GNM\cite{Park2013}, and NMA\cite{Park2013}. These methods cover statistical learning-based approaches, physics-inspired coarse-grained models, and topology-based descriptors, providing a comprehensive benchmark for performance assessment.

Table~\ref{tab:comparison_all} summarizes the results of CAL and all baseline methods across four benchmark datasets (small, medium, large, and superset) in terms of Pearson correlation coefficient (PCC).

\begin{table}[h]
\centering
\caption{Comparison of CAL with existing methods in terms of Pearson correlation coefficient (PCC) on four benchmark datasets.}
\label{tab:comparison_all}
\begin{tabular}{lcccccccc}
\hline
Protein Set & CAL & ASPH (B) & ASPH (W) & opFRI & pfFRI & GNM & NMA \\
\hline
Small    & 0.851 & 0.85 & 0.86 & 0.667 & 0.594 & 0.541 & 0.480 \\
Medium   & 0.777 & 0.69 & 0.69 & 0.664 & 0.605 & 0.550 & 0.482 \\
Large    & 0.715 & 0.61 & 0.62 & 0.636 & 0.591 & 0.529 & 0.494 \\
Superset & 0.760 & 0.65 & 0.66 & 0.673 & 0.626 & 0.565 & NA \\
\hline
\end{tabular}
\end{table}

\subsection{Blind Prediction}

In this section, we evaluate the predictive performance of the proposed feature representation under a blind prediction setting. The 14-dimensional CAL descriptors are combined with the additional features introduced in Section 2.4 to construct the final input representation. Two ensemble learning models, Random Forest (RF) and Gradient Boosting Decision Tree (GBDT), are employed for B-factor regression.

To comprehensively assess model robustness, we conduct experiments under three evaluation protocols, including protein-level 10-fold cross-validation, atom-level 10-fold cross-validation, and leave-one-protein-out (LOPO) validation. All experiments are performed on four benchmark datasets :small, medium, large, and superset.

we first performed atom-level 10-fold cross-validation experiments on four benchmark datasets, including the \textit{small}, \textit{medium}, \textit{large}, and \textit{superset} collections. In this setting, all residues from all proteins within a dataset were pooled together and randomly partitioned into ten folds regardless of their protein identities. During each iteration, nine folds were used for training and the remaining fold was reserved for testing. To improve the robustness of the evaluation and reduce the influence of random partitioning, the entire 10-fold cross-validation procedure was repeated using ten independent random seeds, resulting in a total of one hundred training and testing runs for each model and dataset combination. The final performance was reported as the mean and standard deviation of the Pearson correlation coefficient (PCC) and root mean square error (RMSE) across all runs.

The prediction results are summarized in Table~\ref{tab:atom_cv}. Overall, both machine learning models achieve strong predictive performance on all four benchmark datasets, indicating that the proposed CAL representation combined with the additional features introduced in Section 2.4 provides an informative description of residue flexibility. Among the two models, Random Forest consistently outperforms Gradient Boosting Decision Trees in terms of both PCC and RMSE on all datasets, suggesting that ensemble averaging and bootstrap aggregation are particularly effective for capturing the nonlinear relationship between the proposed features and experimental B-factors.

On the \textit{superset} dataset, which contains the largest structural diversity and serves as the primary benchmark in this study, Random Forest achieves a PCC of $0.8583 \pm 0.0072$ with an RMSE of $5.9848 \pm 0.1666$, while GBDT obtains a PCC of $0.8554 \pm 0.0072$ and an RMSE of $6.3260 \pm 0.1687$. The relatively small standard deviations observed for both models demonstrate strong robustness and stability under different random data partitions.

For the three subset benchmarks, the prediction accuracy generally follows the trend of \textit{small} $>$ \textit{large} $>$ \textit{medium} in terms of PCC values. The \textit{small} dataset yields the highest accuracy, reaching PCC values of $0.8478$ and $0.8417$ for RF and GBDT, respectively, while the \textit{medium} dataset exhibits comparatively lower correlations of $0.7904$ and $0.7339$. This behavior is likely associated with differences in structural complexity and conformational heterogeneity among proteins of different sizes. Nevertheless, both models maintain PCC values above $0.73$ across all datasets, demonstrating the general applicability and robustness of the proposed representation over a wide range of protein structures. Detailed numerical results are provided in Table~\ref{tab:atom_cv}.

\begin{table}[htbp]
\centering
\caption{Atom-level 10-fold cross-validation results on four benchmark datasets using RF and GBDT models.}
\label{tab:atom_cv}
\begin{tabular}{lcccc}
\hline
Dataset & RF PCC & RF RMSE & GBDT PCC & GBDT RMSE \\
\hline
Small &
$0.848 \pm 0.036$ &
$4.816 \pm 1.296$ &
$0.842 \pm 0.039$ &
$5.170 \pm 1.482$ \\

Medium &
$0.790 \pm 0.026$ &
$3.925 \pm 0.226$ &
$0.733 \pm 0.031$ &
$4.566 \pm 0.254$ \\

Large &
$0.801 \pm 0.053$ &
$5.236 \pm 0.780$ &
$0.791 \pm 0.052$ &
$5.597 \pm 0.753$ \\

Superset &
$0.858 \pm 0.007$ &
$5.985 \pm 0.167$ &
$0.855 \pm 0.007$ &
$6.326 \pm 0.169$ \\
\hline
\end{tabular}
\end{table}

In addition to the atom-level evaluation, we further conducted protein-level 10-fold cross-validation to assess the robustness of the proposed CAL representation under a more challenging and realistic setting. In this experiment, all residues belonging to the same protein were kept together, and proteins were partitioned into ten disjoint folds. Similar to the atom-level setting, the entire procedure was repeated across ten independent random seeds, resulting in a total of 100 independent evaluations for each model and dataset.

The results are summarized in Table~\ref{tab:protein_cv}. Overall, compared with the atom-level setting, protein-level prediction is significantly more challenging, as reflected by a clear decrease in both PCC and an increase in RMSE across all datasets. This indicates that generalizing across different protein structures introduces additional difficulty due to larger structural variability and more complex inter-residue correlations.

Among all datasets, the \textit{small} dataset achieves the best performance, with RF reaching a PCC of 0.6303 ± 0.2306 and GBDT achieving 0.6427 ± 0.1934. As dataset size increases, prediction accuracy gradually decreases. On the \textit{medium} dataset, PCC values drop to approximately 0.48–0.51, while the \textit{large} dataset exhibits further degradation in performance, particularly for RF. This trend suggests that increased structural diversity and conformational heterogeneity negatively affect model generalization.

On the \textit{superset} dataset, which contains the most diverse protein structures, both models achieve moderate predictive performance, with GBDT slightly outperforming RF (0.4561 ± 0.1612 vs. 0.4347 ± 0.1562 in PCC). Although the absolute performance is lower compared to atom-level results, the relatively consistent standard deviations indicate that the proposed feature representation still maintains a certain degree of robustness across different random splits.

Overall, while protein-level prediction is inherently more difficult than residue-level regression, the proposed CAL representation combined with additional features remains effective in capturing global structural trends across proteins of varying sizes.

\begin{table}[htbp]
\centering
\caption{Protein-level 10-fold cross-validation results on four benchmark datasets using RF and GBDT models.}
\label{tab:protein_cv}
\begin{tabular}{lcccc}
\hline
Dataset & RF PCC & RF RMSE & GBDT PCC & GBDT RMSE \\
\hline
Small &
$0.630 \pm 0.231$ &
$6.624 \pm 3.091$ &
$0.643 \pm 0.193$ &
$6.490 \pm 3.094$ \\

Medium &
$0.484 \pm 0.174$ &
$5.721 \pm 0.975$ &
$0.512 \pm 0.138$ &
$5.560 \pm 1.050$ \\

Large &
$0.412 \pm 0.224$ &
$7.978 \pm 2.362$ &
$0.514 \pm 0.145$ &
$7.321 \pm 2.337$ \\

Superset &
$0.435 \pm 0.156$ &
$10.151 \pm 5.955$ &
$0.456 \pm 0.161$ &
$9.716 \pm 5.551$ \\
\hline
\end{tabular}
\end{table}

We further evaluated the proposed CAL representation under the leave-one-protein-out (LOPO) setting across the four benchmark datasets. In this setting, each protein is sequentially used as an independent test case, while all remaining proteins are used for training. This protocol provides a rigorous evaluation of model performance under true cross-protein generalization, where no structural information from the test protein is available during training.

The results are summarized in Table~\ref{tab:lopo}. Overall, compared with both atom-level and protein-level cross-validation settings, LOPO exhibits a noticeable decline in predictive performance, indicating the significantly increased difficulty of generalizing to completely unseen protein structures. This further highlights the challenge of capturing transferable structural patterns across diverse protein conformations.

Among all datasets, the \textit{medium} and \textit{superset} datasets achieve relatively better performance compared to the \textit{small} and \textit{large} subsets. In particular, GBDT consistently outperforms RF across all datasets, suggesting that gradient boosting is more effective in handling the high variance introduced by extreme cross-protein heterogeneity.

On the \textit{superset} dataset, which contains the largest number of proteins (345 in total), GBDT achieves a PCC of 0.6054 ± 0.1789 with an RMSE of 5.9954 ± 3.6669, while RF obtains a PCC of 0.5686 ± 0.2177 and an RMSE of 6.1160 ± 3.8282. Despite the increased variance, these results demonstrate that the proposed feature representation still preserves a moderate level of predictive capability even under the most challenging LOPO setting.

Overall, although LOPO significantly increases the difficulty of prediction, the CAL representation combined with additional features remains capable of capturing transferable structural signals across unseen proteins.

\begin{table}[htbp]
\centering
\caption{Leave-one-protein-out cross-validation results on four benchmark datasets using RF and GBDT models.}
\label{tab:lopo}
\begin{tabular}{lcccc}
\hline
Dataset & RF PCC & RF RMSE & GBDT PCC & GBDT RMSE \\
\hline
Small &
$0.412 \pm 0.341$ &
$5.161 \pm 4.304$ &
$0.527 \pm 0.274$ &
$5.247 \pm 3.935$ \\

Medium &
$0.567 \pm 0.184$ &
$5.543 \pm 1.681$ &
$0.611 \pm 0.163$ &
$5.352 \pm 1.937$ \\

Large &
$0.513 \pm 0.180$ &
$7.077 \pm 3.431$ &
$0.562 \pm 0.139$ &
$6.734 \pm 3.272$ \\

Superset &
$0.569 \pm 0.218$ &
$6.116 \pm 3.828$ &
$0.605 \pm 0.179$ &
$5.995 \pm 3.667$ \\
\hline
\end{tabular}
\end{table}

\subsection{Protein Flexibility Case Studies}

To further evaluate the proposed method from a structural perspective, four representative proteins (1CLL, 2GZQ, 1V70, and 2HKQ) were selected for detailed case studies. For each protein, three-dimensional flexibility distributions and residue-level B-factor predictions were analyzed to assess the ability of the proposed method to capture local flexibility patterns in comparison with the experimental B-factors and the Gaussian Network Model (GNM).

Protein structures were visualized using Visual Molecular Dynamics (VMD)\cite{Humphrey1996}. The residue color gradually changes from blue to red, representing increasing flexibility (higher B-factors), whereas blue regions correspond to relatively rigid structural regions. Detailed interpretations of the structural characteristics and prediction results for each protein are presented in the following subsections.

1CLL corresponds to calmodulin, a highly conserved calcium-binding protein that plays an essential role in intracellular calcium signaling. Its structure consists of two globular domains connected by a long central $\alpha$-helical linker, resulting in distinct regional flexibility.

The experimental B-factor profile indicates that the central $\alpha$-helical linker is the most flexible region of the protein, whereas the two terminal domains remain comparatively rigid. CAL successfully reproduces this flexibility distribution and accurately captures the pronounced local flexibility peak in the linker region between residues 75 and 81. In contrast, GNM method predicts this entire region as an almost uniformly low-flexibility segment, substantially underestimating the mobility of the linker and producing the largest discrepancy with the experimental observations. These results demonstrate that CAL effectively captures localized flexibility variations within the linker region, whereas the traditional GNM method fails to describe such regional flexibility fluctuations.

\begin{figure}[H]
    \centering
    \includegraphics[width=0.9\linewidth]{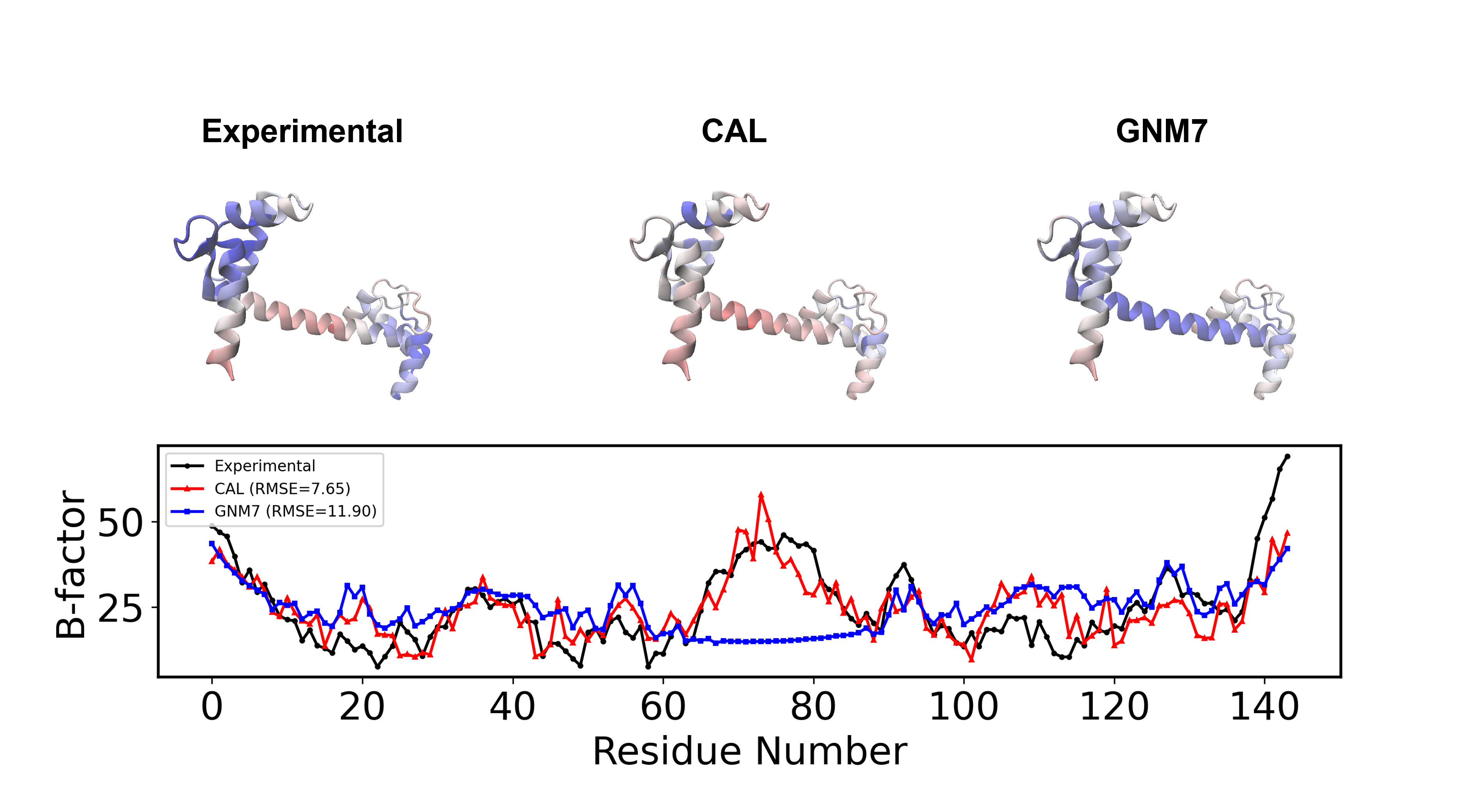}
    \caption{
    Comparison of flexibility prediction results for protein 1CLL. The top row shows the three-dimensional visualization of the experimental B-factor, CAL prediction, and GNM method prediction mapped onto the protein structure, where blue indicates relatively rigid regions and red indicates relatively flexible regions. The bottom panel presents the residue-wise B-factor profiles, with the experimental values shown in black, CAL predictions in red, and GNM method predictions in blue. CAL achieves a PCC of 0.748 and an RMSE of 7.65, compared with a PCC of 0.232 and an RMSE of 11.90 obtained by GNM method.
    }
    \label{fig:1cll}
\end{figure}

Beyond the central linker, CAL also provides a more faithful description of the global flexibility pattern, successfully following the experimental variations in both the N-terminal and C-terminal regions and reproducing the increasing flexibility toward the C-terminus. Compared with GNM method, CAL reduces the RMSE from 11.90 to 7.65, corresponding to an error reduction of approximately 36\%. Meanwhile, the PCC improves from 0.232 to 0.748. These results indicate that CAL not only preserves the overall flexibility distribution but also more accurately characterizes localized flexibility peaks and regional flexibility differences. This case study is further illustrated in \autoref{fig:1cll}.

\begin{figure}[H]
    \centering
    \includegraphics[width=0.9\linewidth]{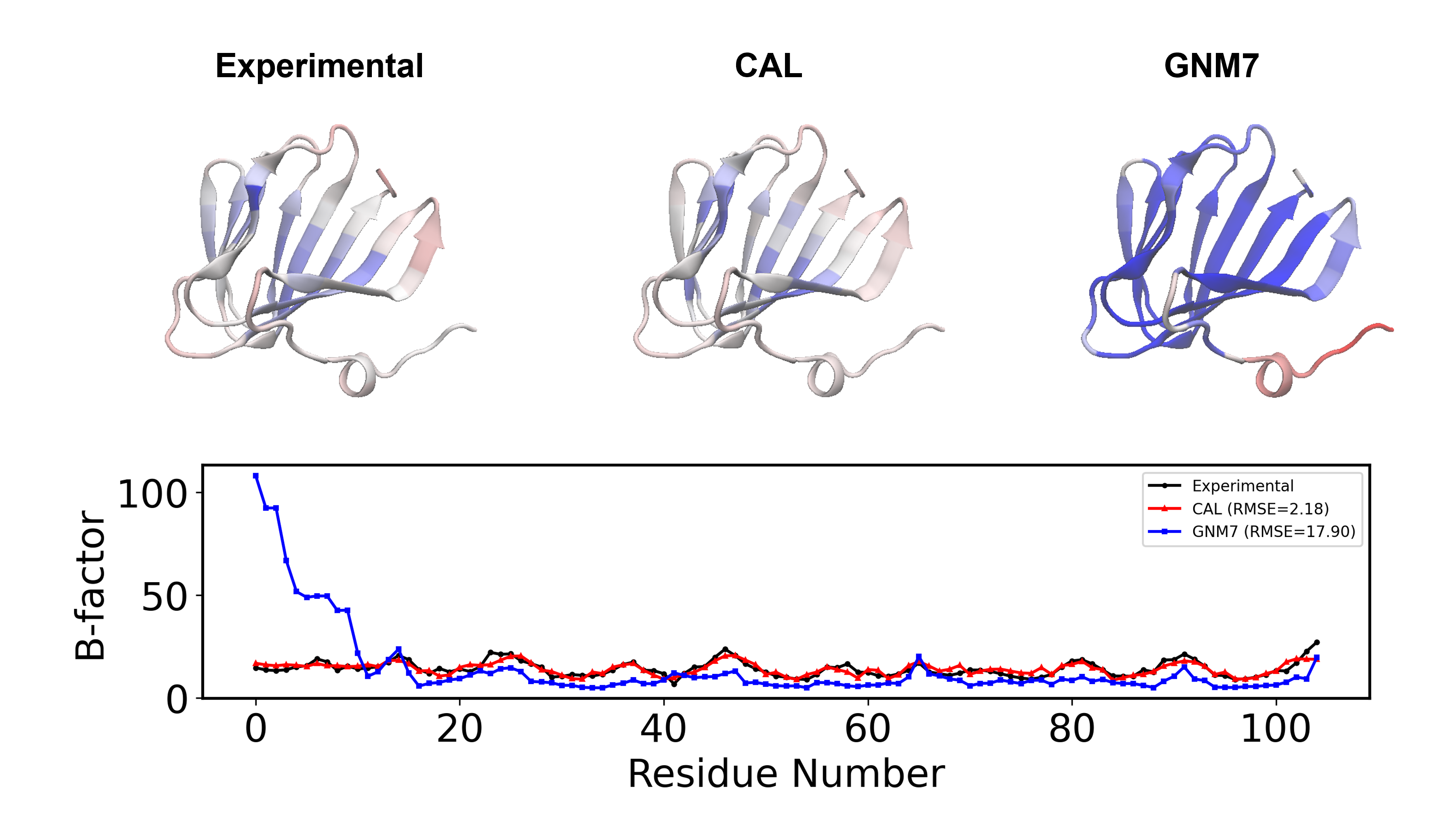}
    \caption{
    Comparison of flexibility prediction results for the protein with PDB ID 1V70. CAL shows strong agreement with experimental B-factors, achieving a PCC of 0.8076 and an RMSE of 2.18, while GNM method yields a PCC of 0.1648 and an RMSE of 17.52, demonstrating significantly larger prediction errors.
    }
    \label{fig:1v70}
\end{figure}

We further analyze the protein with PDB ID 1V70, which exhibits a heterogeneous experimental B-factor distribution characterized by higher flexibility at the N-terminal region, relatively stable central segments, and moderate local fluctuations along the sequence. Quantitatively, CAL shows strong agreement with experimental data, achieving a Pearson correlation coefficient of 0.8076 and an RMSE of 2.18, indicating that it accurately captures both the global flexibility trend and local structural variations while maintaining amplitude consistency with the experimental profile. In contrast, GNM method deviates substantially from the experimental distribution, with a PCC of 0.1648 and an RMSE of 17.52, reflecting significant predictive errors. The error is primarily concentrated in the N-terminal region, where systematic overestimation dominates and accumulates into the largest contribution to the overall deviation; this behavior is consistent with previous reports indicating that GNM method tends to overestimate flexibility in low residue-index regions, thereby amplifying overall RMSE\cite{Hayes2025}. Overall, CAL demonstrates stable and accurate predictive performance for this protein, successfully reproducing both global trends and local flexibility variations. This case study is further illustrated in \autoref{fig:1v70}.

\begin{figure}[H]
    \centering
    \includegraphics[width=0.9\linewidth]{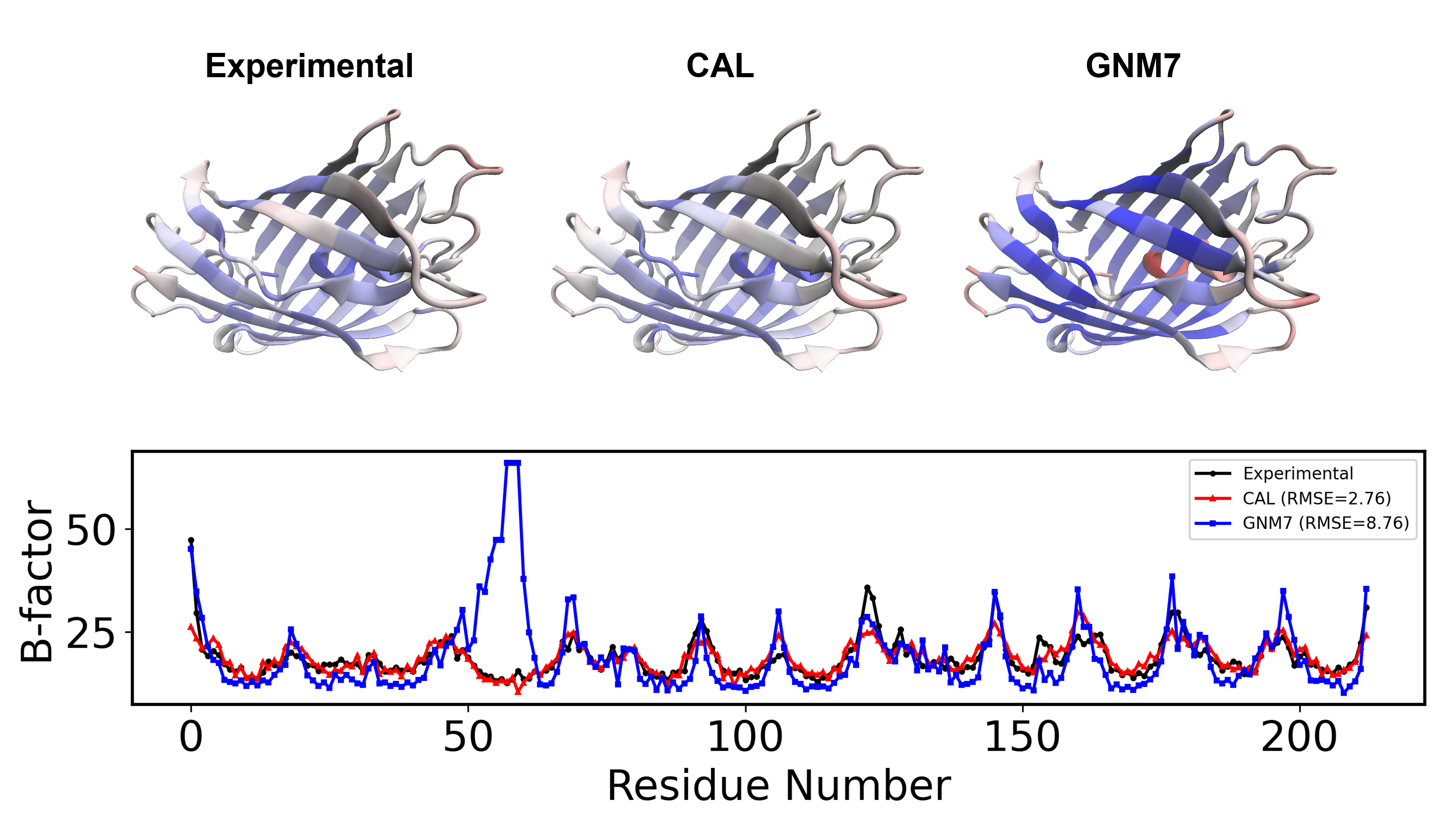}
    \caption{
    Comparison of flexibility prediction results for the protein with PDB ID 2HKQ. The top row shows the structural mapping of experimental B-factors, CAL predictions, and GNM method predictions, while the bottom panel presents the residue-wise B-factor profiles. Quantitatively, CAL achieves a PCC of 0.7940 and an RMSE of 2.76, whereas GNM method yields a PCC of 0.3611 and an RMSE of 8.76.
    }
    \label{fig:2hqk}
\end{figure}

We next present the flexibility prediction results for the protein with PDB ID 2HKQ. 2HKQ is a fluorescence-related protein widely used in bioimaging and molecular labeling applications. Its overall structure is dominated by $\beta$-sheet architecture, while certain loop and terminal regions still exhibit moderate local flexibility. From a global perspective, the experimental B-factor has a mean value of 18.68, which is almost identical to the CAL prediction mean of 18.68. In addition, the standard deviation decreases from 4.54 in the experimental profile to 3.60 in CAL predictions, indicating that CAL preserves the global amplitude distribution while introducing mild smoothing that enhances structural consistency. In terms of predictive performance, CAL achieves a Pearson correlation coefficient of 0.7940 and an RMSE of 2.76, accurately reproducing both global trends and local fluctuations in the experimental B-factor distribution. In contrast, GNM method performs significantly worse, with a PCC of 0.3611 and an RMSE of 8.76. Its standard deviation increases to 9.31, indicating substantial over-fluctuation and instability in the predicted profile. Error analysis further reveals that GNM method exhibits pronounced local deviations around residues 63–65, where error peaks exceed 50, making this region the dominant contributor to the overall RMSE. This suggests a systematic overestimation of flexibility in locally perturbed regions. In comparison, CAL maintains consistently low errors in this region and exhibits a more uniform error distribution across the sequence, demonstrating stronger robustness and structural consistency. Overall, CAL not only agrees well with global statistical properties but also avoids the pronounced local deviations observed in GNM method, providing a more stable and physically consistent prediction of residue-level flexibility. This case study is further illustrated in ~\autoref{fig:2hqk}.

Finally, we present the flexibility prediction results for the protein with PDB ID 2GZQ. The protein exhibits moderately varying B-factor distributions, with an experimental mean of 5.12 and a standard deviation of 3.83, indicating a relatively balanced flexibility profile with localized fluctuations.

To further investigate the contribution of larger neighborhood information to flexibility prediction, an additional spatial scale with $R=25$ was incorporated into the CAL representation. This extension enriches the original multiscale description and increases the feature dimension from 14 to 22 through the inclusion of an additional set of topological descriptors.

With the enriched representation, CAL achieves a Pearson correlation coefficient (PCC) of 0.4689 and an RMSE of 3.37, outperforming the GNM method, which obtains a PCC of 0.3972 and an RMSE of 3.55. The improvement in both correlation and prediction error suggests that the additional spatial scale provides complementary structural information that was not fully captured by the original feature set.

From a local perspective, CAL reproduces the experimental fluctuations more accurately across several moderately flexible regions while avoiding excessive amplification of individual peaks. The predicted standard deviation of CAL is 1.80, compared to 2.09 for GNM, indicating that CAL maintains smoother and more stable predictions while preserving the overall flexibility pattern of the protein.

\begin{figure}[H]
    \centering
    \includegraphics[width=0.9\linewidth]{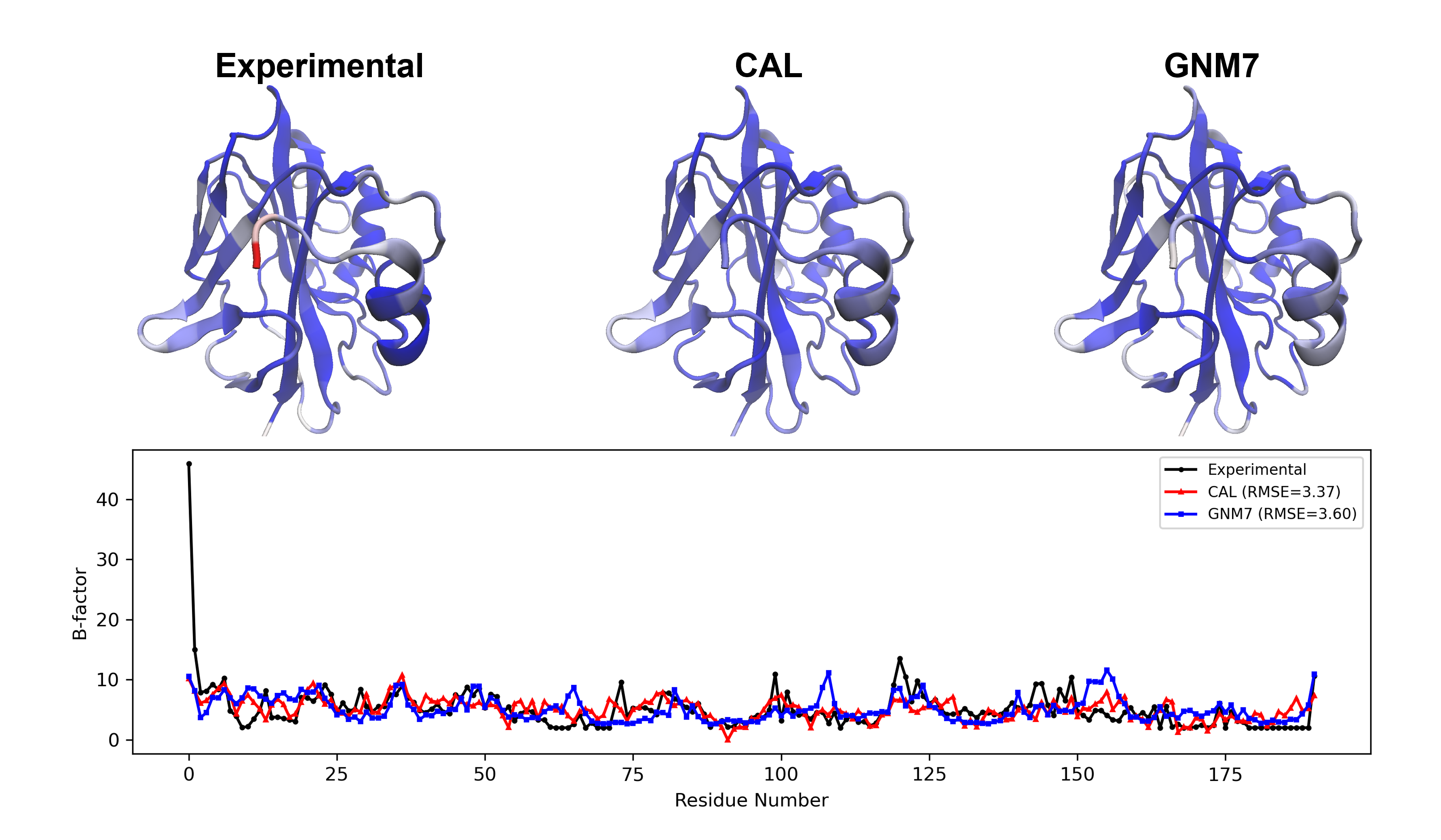}
    \caption{
    Comparison of flexibility prediction results for PDB 2GZQ. The top row shows the structural mapping of experimental B-factors, CAL predictions, and GNM7 predictions, while the bottom panel presents the residue-wise B-factor profiles. Quantitatively, CAL achieves a PCC of 0.4689 and an RMSE of 3.37, outperforming GNM7, which obtains a PCC of 0.3972 and an RMSE of 3.60. Although both methods capture the overall flexibility pattern of the protein, CAL provides improved agreement with experimental measurements and exhibits smoother local variations across residue positions.
    }
    \label{fig:2gzq}
\end{figure}

Overall, 2GZQ demonstrates the importance of incorporating additional multiscale information for proteins with relatively complex flexibility distributions. The expanded 22-dimensional CAL representation significantly improves predictive performance and further highlights the ability of commutative algebra-based descriptors to capture nontrivial structural organization across multiple spatial scales. This case study is further illustrated in \autoref{fig:2gzq}.

\section{Conclusion}\label{sec:conclusion}

In this work, we propose a novel structural representation method for protein flexibility prediction, termed Commutative Algebra Learning (CAL). The method constructs multiscale combinatorial descriptors based on Vietoris--Rips filtrations and f-vector statistics, enabling a compact yet expressive representation of local protein geometry. Unlike traditional physics-based models that rely on explicit dynamical simulations or matrix diagonalization, CAL directly encodes structural information through combinatorial evolution across spatial scales.

Extensive experiments on four benchmark datasets demonstrate the effectiveness and robustness of the proposed method. Under atom-level and protein-level 10-fold cross-validation, CAL combined with machine learning models (RF and GBDT) consistently achieves strong predictive performance, outperforming or remaining competitive with several representative baseline methods, including PSL, ASPH, opFRI, pfFRI, GNM, and NMA. Furthermore, in the more challenging leave-one-protein-out setting, CAL maintains stable predictive capability across diverse protein structures, although with an expected performance drop due to significantly increased generalization difficulty.

Overall, the results indicate that CAL provides a powerful and flexible framework for protein flexibility analysis. By integrating multiscale topological information with additional structural descriptors, the proposed approach effectively captures both local and global structural variations, offering a promising direction for data-driven modeling of protein dynamics.

		\section{Code and Data availability}
		
 All data and the code needed to reproduce this paper's result can be found at\\ \href{https://github.com/ZHHMCI/CAL}{https://github.com/ZHHMCI/CAL}.

\section{Acknowledgments}
This work was supported in part by University of North Carolina at Charlotte KCOS Faculty Research Grant.

\bibliographystyle{unsrt}

\end{document}